\documentclass[a4paper,8pt,final,onecolumn]{article}
\usepackage{graphicx}
\newlength{\vshift}
\input epsf.tex
\newlength{\hshift}

\usepackage{amssymb}
\usepackage{amsmath}
\usepackage{amsthm}
\usepackage{amsopn}
\usepackage{pifont}
\usepackage[normalem]{ulem}
\usepackage{xcolor}
\usepackage{cite}
\usepackage[pagebackref=true,colorlinks,linkcolor=blue,citecolor=red]{hyperref}
\def\uno{\mbox{1 \kern-.59em {\rm l}}}

\def\ba{\begin{eqnarray}}
\def\ea{\end{eqnarray}}
\def\la{\langle}
\def\ra{\rangle}

\def\hh{\hskip 2cm}
\def\lo{\longrightarrow}

\def\be{\begin{equation}}
\def\ee{\end{equation}}
\def\bea{\begin{eqnarray}}
\def\eea{\end{eqnarray}}
\def\lo{\longrightarrow}

\def\hh{\hskip 2cm}
\def\la{\langle}
\def\ra{\rangle}

\def\m{\mu}
\def\n{\nu}

\begin{document}

\vspace*{1cm}

\begin{center}

{\bf \large The quantum (non-Abelian) Potts model and its exact solution}

\vskip 4em

{\bf  Razieh Mohseninia\footnote{Email: mohseninia@physics.sharif.ir}}  \  and
{\bf Vahid ~Karimipour\footnote{Email: vahid@sharif.edu}} 

\vskip 1em
Department of Physics,\\  Sharif University of Technology,
P.O. Box 111555-9161,
Tehran, Iran.\\

\end{center}

\vspace*{1.9cm}

\begin{abstract}
We generalize the classical one dimensional Potts model to the case where the symmetry group is a non-Abelian finite group. It turns out that this new model has a quantum nature in that its spectrum of energy eigenstates consists of entangled states.  We determine the complete energy spectrum, i.e. the ground states and all the excited states with their degeneracy structure. We calculate the  partition function by two different algebraic and combinatorial methods. We also determine the entanglement properties of its ground states.  
\end{abstract}

\section{Introduction}\label{intro}

The Ising model is a prototype of a statistical mechanical model for studying order-disorder transitions. It is also the first statistical model which has led to exact solution in one and two dimensions.  Since its inception\cite{Ising} this model has triggered an intense effort in investigation of other models resulting in an extensive literature on the subject \cite{lit1,lit2,lit3,lit4,lit5,lit6,lit7,lit8,lit9,potts}. There are now a  large library of  statistical and quantum mechanical models, differing in their degrees of freedom, interaction type, the type and dimensions of lattices,  and of course methods of solution \cite{exact1,exact2,model1,model2,model3}.  There are also models which can be called integrable, meaning that they allow sufficient number of conserved quantities, leading to a full determination of their spectra and other observable quantities. Among the well-known classical models, we can mention particularly the $d$-state Potts model \cite{potts}, which is the simplest generalization of the Ising model in that the degrees of freedom take $d$ instead of two different values and  interact according to the Hamiltonian 
\be\label{Hpot}
H=-\sum_{\la i,j\ra} \delta(s_i, s_j),
\ee
where $\la , \ra $ means that interactions are between nearest neighbors and $s_i$ takes $d$ different values. This is a classical model in that every configuration of the so-called spin variables $s_i$ is an energy eigenstate.  The importance of this model, like Ising model, is that it can be mapped to many other important models in science, i.e. the famous four-color problem in the case of Potts model, \cite{app1,app2,app3,app4}.  The Ising model has a $Z_2$ symmetry, meaning that the configurations $(s_1, s_2, \cdots s_n)$ and $(-s_1, -s_2, \cdots -s_N)$ both have the same energy. ]In the Potts model, this symmetry has been elevated to the $Z_d$ group, where now due to the form of the Hamiltonian (\ref{Hpot}), one can shift all the spin variables by any integer value $k \in [0,\cdots d-1]$  and the energy remains the same. This symmetry can be seen in a better way if we use an identity and write the local energy term as  
\be
\delta(s,s')=\frac{1}{d}\sum_{n=0}^{d-1} \omega^{ns}\omega^{-ns'},
\ee
where $\omega$ is a $d-$th root of unity. \\

It is the purpose of this work to generalize the Potts and indeed Ising model to the case where the symmetry group is a finite  (Non-Abelian) group. Such a model, if properly defined will certainly have a very rich structure and will certainly deepen and widen our perspective of exactly solvable models of statistical mechanics and integrable models. Moreover it has the potential of enriching our knowledge of order-disorder transitions in statistical mechanics.  Of course the model as defined now is mainly of theoretical interest and it may be difficult to make concrete connections with specific physical problems. We only hope with hindsight and in view of the experience with other more or less abstract models in statistical mechanics like the face models and vertex models \cite{exact1}, this new model will also find a proper place in the library of exactly solvable models, possibly with new applications in the future. \\

We will focus our attention to a one dimensional lattice and show that such a natural generalization is indeed possible and will define the non-Abelian Potts model. We determine the full spectrum and show that the ground states and indeed many of the excited states of this model are entangled. We will determine the amount of this entanglement both for a single site and for a block of finite length. We also determine the entanglement between two different sites of the lattice.  In this sense the model is a quantum mechanical model, in contrast to classical models where there is no entanglement in their energy eigenstates. It turns our that the properties of irredudcible representations play an important role in  the nature of the spectrum and its entanglement properties.
We will also calculate the partition function of the model in two different ways, that is  we follow an algebraic approach where we calculate the trace of the thermal state and a combinatorial approach where we also count the degeneracy of all energy levels. The results of the two approaches, which correspond respectively to the high and low temperature expansions,  agree as they should.  In all aspects the results pertaining to this model reduce to the Potts model when the symmetry group $G$ reduces to the Abelian group $Z_d$. \\

The structure of this paper is as follows: In section \ref{pre} we review the preliminary material from theory of finite groups and their representations. In section \ref{model}, we define the non-Abelian model on a one dimensional lattice, which as we see, is  a special case of the Kitaev quantum double model \cite{kitaev}. In two dimensions this correspondence in no longer true, since as is well known the Kitaev model entails four-body interactions while our model entails two-body interactions, as it should as a generalization of Potts model. In fact generalization of the Potts model to two dimensions is a non-trivial problem as we will discuss in the discussion.  In section \ref{spec} we derive the ground states and determine their entanglement properties, and in section \ref{excited states} we determine all the excited states, hence the full spectrum. In section \ref{part}, we calculate the partition function in closed form, by summing up the terms in both high and low temperature expansions. In all our study we make contact with the simpler and special case where the group we are considering is the Abelian group $Z_d$, and the model is the d-State Potts model.  
 
\section{Preliminaries}\label{pre}
Let $G$ be a finite group of size $|G|$. 
 Two elements $a$ and $b$ of the group $G$ are said to be conjugate if there is a $g\in G$, such that $a=gbg^{-1}$. This is an equivalence relation which partitions the group into conjugacy classes. The conjugacy class of an element $a$ is denoted by $C_a$. The number of conjugacy classes is denoted by $K$. This equals the number of irreducible representations of the group. It is customary to denote the former by Latin indices and the latter by Greek indices, hence we have $C_i$ as a conjugacy class and $D^\m$ as an Irreducible representation.  The number of elements in $C_i$ is denoted by $|C_i|$, and the dimension of an irreducible representation $D^\m$ is denoted by $n_\m$. We then have $\sum_{i=1}^K |C_i|=|G|$. Let $V$ be a vector space spanned by orthonormal vectors $|g\ra \ g\in G$. On this vector space, two regular representations called left and right actions respectively, are defined as 

\be
L(g) |h\ra = |gh\ra, \qquad \qquad R(g)|h\ra=|hg^{-1}\ra.
\ee
Obviously the two kinds of actions commute with each other. For an abelian group, $[L(g), L(g')]=[R(g),R(g')]=0$ leading to a simple disentangled spectrum as we will see in sections \ref{model}, while for a non-abelian group this is no longer the case. \\

Both regular representations decompose into a sum of irreducible representations, each representation $D^\m$ occurring with a multiplicity equal to its dimension $n_\m$, hence we have the relation
$
\sum_{\m=1}^K n_\m^2 = |G|.
$

\begin{figure}[t]
\begin{center}
\includegraphics[scale=0.6]{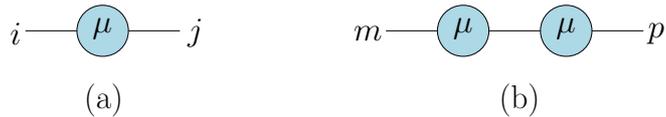}
\caption{(Color online) Graphical representation for $|D^\m_{ij}\ra$'s and their contraction: a) a representation for $|D^\m_{ij}\ra$. b) a representation for $\sum_n|D^\m_{mn}\ra |D^\m_{np}\ra$.}
\label{contraction}
\end{center}
\end{figure}

 The matrix entries  of an irreducible representation $D^{\mu}$ of an element $g\in G$ are denoted by $D^\mu_{mn}(g)$.  
 For each set of labels $\mu , m, n$, a $|G|$ dimensional vector  $|D^\m_{m,n}\ra\in { V}$ is defined as:
\be \label{basis}
|D_{mn}^\mu \ra := \sqrt{\frac{n_\mu}{|G|}}  \sum_{ g\in G} D^\mu_{mn}(g) |g\ra.
\ee
This is in fact the Fourier transform on the finite group $G$. It is well known that these vectors provide an orthonormal and complete basis for ${V}$ \cite{grouptheory}. The orthonormality implies that 

\be \label{orthonormal}
\la D^\m_{m,n}|D^\n_{p,q}\ra=\delta^{\m,\n}\delta_{m,p}\delta_{n,q}, 
\ee
or   
\be \label{orthonormality}
\sum_{g \in G} D^\mu_{mi}(g) \overline{D^{\nu}_{nj}(g)} =\frac{|G|}{n_\mu} \delta^{\mu \nu} \delta_{mn} \delta_{ij}.
\ee Here $\overline{x}$ denotes complex conjugate of $x$. We will later use a graphical representation for $|D^\m_{ij}\ra$ which we depict in figure \ref{contraction}. \\

For each representation $D^\m$, the character of a conjugacy class $C_i$, is defined as $\chi^\m_i:=\text{Tr}(D^\m(g))$, where $g\in C_i$. The significance of characters of irreducible representations  is that they are orthonormal and complete in the sense that if we define the $K$ dimensional formal vectors 
$|\chi^\m\ra:=\frac{1}{\sqrt{|G|}}\sum_{i=1}^K \sqrt{|C_i|} \chi^\m_i |i\ra$, where $|i\ra$ 's form a computational basis, then these new vectors form a complete orthonormal basis. The orthonormality implies that
\be \label{ccort}
\sum_{g \in G} \chi^\m(g)\overline{\chi^\n(g)}=|G| \delta^{\m,\n}.
\ee

Finally note that the product of irreducible representations decomposes into sum of them in the form 
\be
D^\m(g) \otimes D^\n(g) = \bigoplus_\lambda F^{\m,\n}_\lambda D^\lambda(g),
\ee
where $F^{\m,\n}_\lambda$ are the (Clebsh-Gordan) or fusion coefficients of this group. Taking the trace of both sides, this leads to 
\be\label{chichi}
\chi^\m(g) \chi^\n(g) = \sum_\lambda F^{\m,\n}_\lambda \chi^\lambda(g).
\ee
 
We are now equipped with almost all the necessary ingredients of group theory to define and study the non-Abelian Potts model.  
\section{ The Model}\label{model}

Consider a periodic chain of length $N$.  Let $G$ be a finite (Abelian or non-Abelian) group of size $|G|$. To each site of the chain, a $|G|$ dimensional Hilbert space $V$ is assigned with orthonormal basis vectors  $\lbrace |g\ra, g \in G \rbrace$. Therefore we have $\la g|g'\ra= \delta_{g,g'}$ and $\sum_g|g\ra\la g|=I_ {V}$. The dimensional of the full Hilbert space ${\cal V}:=V^{\otimes N}$ is given by $|G|^N$. \\

The Hamiltonian of the model is the reduction of the quantum double Hamiltonian \cite{kitaev}, in which only the vertex operators are retained and all the plaquette operators are ignored.  Since in a one-dimensional lattice, there is a one to one correspondence between vertices and links, we have used this liberty to put the dynamical (generalized spin) variables on the vertices instead of the links. This makes the model much more akin to the other models of statistical mechanics, like the Ising, Potts and Heisenberg models. To each pair of neighboring sites $(i,\ i+1)$, we assign the following local Hamiltonian
\be \label{localH}
H_i=\frac{1}{|G|}\sum_{g\in G} R_{i}(g) L_{i+1}(g),
\ee 
where the indices $i$ and $i+1$ show that $R_i$ and $L_{i+1}$ act on these two sites respectively. It is easily verified that $H_i$ is a projector, namely $H_i^2=H_i$ and $[H_i,\ H_j]=0,\ \ \forall \ \ i, j$. Therefore the eigenvalues of $H_i$ are restricted to $0$ and $1$. The total Hamiltonian is the sum of all these local Hamiltonians, i.e. 

\be \label{hamiltonian}
H=- \sum_{i=1}^N H_{i}, 
\ee
where periodic boundary condition $N+1\equiv 1$ is implied in all states and operators. \\

Consider the special case where $G$ is an Abelian group, and in particular $Z_d$  the finite cyclic group of order $d$, 
$
Z_d :\lbrace 0,1, 2, 3, ... , d-1 \rbrace,
$
where the group operation is addition modulo $d$. Then we have \\
\be
L(g) |h\ra= |h+g\ra, \qquad \qquad R(g)|h\ra=|h-g\ra \ \ \ \ \ \ \ \ \ \  h,\ g\in Z_d,\\ \nonumber
\ee
leading to 
\be
L(g) \equiv X^g, \qquad \qquad \qquad R(g)\equiv X^{-g},
\ee
where $X=\sum_g |g+1\ra\la g|$ is the shift operator or the $d-$ dimensional generalization of the Pauli matrix $\sigma_x$, and $X^g$ is $X$ to the power $g$. Therefore in the $Z_d$ case, the local Hamiltonian \ref{localH} can be written as 
\be
H_i= \frac{1}{d} \sum_{g=0}^{d-1} X_i^{-g} X_{i+1}^{g}.
\ee
To see the relevance of this local Hamiltonian  to the Potts model, we note that in this Abelian case and only in this case, all the operators $L(g)\equiv X^g$ can be diagonalized in the same basis, since $[L(g), L(g')]=0$ for an Abelian group. This is the basis in which the shift operator $X$ is diagonal. This operator has the property $X^d=I$ which means that its eigenvalues are $\omega^s,\ \ \ s=0, \ 1, \ \cdots \ d-1$, where $\omega$ is the $d-$th root of unity, $\omega:=e^{\frac{2\pi i}{d}}$. The eigenvectors of $X$ are easily found

\be \label{Dstates}
|D^s\ra:=\frac{1}{\sqrt{d}}\sum_{j=0}^{d-1} \omega^{-js}|j\ra,\hh X|D^s\ra=\omega^s|D^s\ra.
\ee
Now any product state of the form 
\be
|\Psi^{{\bf s}}\ra:=|D^{s_1}\ra\otimes |D^{s_2}\ra\otimes\cdots |D^{s_{i}}\ra\otimes|D^{s_{i+1}}\ra\otimes \cdots |D^{s_N}\ra,
\ee
is an eigestate of all $H_i$'s  and hence the total $H$. In such a basis, $H_i$ can be replaced with its eigenvalues.
Then we have 
\ba
H_i&\lo & \frac{1}{d}\sum_{g=0}^{d-1}\omega^{(s_{i+1}-s_i)g}= \delta_{s_i,s_{i+1}}.
\ea
Hence in the special case $G=Z_d$, we will have the classical $d-$state Potts model described by 
\be \label{Pottsdelta}
H=-\sum_{i=1}^N \delta_{s_i, s_{i+1}},
\ee
where $s_i$'s take $d$ different values. We now turn to the spectrum when $G$ is a finite non-Abelian group. First we find the ground states and then determine all the excited states.  \\

\section{The ground states}\label{spec}
In order to find the ground states of the Hamiltonian \ref{hamiltonian}, we first consider the Potts model which will act as a guideline for the more complicated non-Abelian case. As equation \ref{Pottsdelta} indicates, the ground state of the Potts model is when all the labels $s_i$ are equal to each other, so there are $d$ different ground states, namely

\be
|\Psi^s\ra:=|D^{s}\ra\otimes |D^{s}\ra\otimes\cdots  |D^{s}\ra.
\ee
In view of the explicit form of the states $|D^s\ra$ in \ref{Dstates}, we can rewrite this as 
\be
|\Psi^s\ra=\frac{1}{\sqrt{d^N}}\sum_{j_1, j_2, \cdots j_N}\omega^{-j_1s}\omega^{-j_2s}\cdots \omega^{-j_Ns}|j_1, j_2,\cdots j_N\ra.
\ee
This can be written as a Matrix Product State (MPS) as 
\be
|\Psi^s\ra=\frac{1}{\sqrt{d^N}}\sum_{j_1, j_2, \cdots j_N}\psi^s(j_1, j_2, \cdots j_N)|j_1, j_2,\cdots j_N\ra,
\ee
where
\be \label{MPSPotts}
\psi^s(j_1, j_2, \cdots j_N)= \text{Tr}(D^{s}(j_1)D^{s}(j_2)\cdots D^{s}(j_N)),
\ee
in which $D^s(j):=\omega^{-sj}$ is the $D^s$ representation of element $j\in Z_d$. \\

We now go on to consider the spectrum of the non-Abelian case. 
Motivated by the MPS representation of the Abelian (Potts) model in \ref{MPSPotts}, we form the following states:
\ba
|\Psi^\mu\ra&:=&\frac{1}{\sqrt{Z}} \text{Tr}(D^{\m}(g_1)D^{\m}(g_2)\cdots D^{\m}(g_N))|g_1, g_2, \cdots g_N\ra\cr &=&\frac{1}{\sqrt{Z}}\chi^\m(g_1g_2\cdots g_N)|g_1, g_2, \cdots g_N\ra,
\ea
where a sum over all group elements $g_i$ is implied. 
Here $Z$ is a normalization factor and $D^\m(g)$ is the matrix of $g$ in the irreducible representation $D^\m$. It is then obvious that these states are eigenstates of the operators $H_i$. To see this we note that  
\ba
H_i|\Psi^\m\ra&=& \frac{1}{|G|}\sum_{g\in G}R_{i}(g)L_{i+1}(g)|\Psi^\m\ra \cr &=& \frac{1}{|G|\sqrt{Z}}\sum_{g\in G}\chi^\m(g_1g_2\cdots g_N)|g_1, g_2 \cdots g_{i}g^{-1}, gg_{i+1}, \cdots g_N\ra=|\Psi^\m\ra,
\ea
where in the last line we have relabeled the group elements  $g_{i}g^{-1}$ to $g'_{i}$ and $gg_{i+1}$ to $g'_{i+1}$ and have used the fact that a sum over all group elements is performed. Therefore for each irreducible representation of the group we find one ground state. Thus the ground state is $K$-fold degenerate, where $K$ is the number of in-equivalent irreducible representations, or equivalently the number of conjugacy classes. To find the normalization, we use the orthogonality relation of characters which reads from \ref{ccort} $\frac{1}{|G|}\sum_{i=1}^K |C_i| |\chi^\m_i|^2=1$ or  $\frac{1}{|G|}\sum_{g\in G}|\chi^\m(g)|^2=1$. Therefore we find
\be
\la \Psi^\m|\Psi^\m\ra=\frac{1}{Z}\sum_{g_1, \cdots g_N} |\chi(g_1\cdots g_N)|^2=\frac{1}{Z}|G|^{N-1}\sum_{g\in G}|\chi(g)|^2=\frac{1}{Z}|G|^N,
\ee
giving the normalized (in fact orthogonal) ground states 
\be \label{gs}
|\Psi^\mu\ra=\frac{1}{\sqrt{|G|^N}}\chi^\m(g_1g_2\cdots g_N)|g_1, g_2, \cdots g_N\ra.
\ee
Note that in contrast to the Abelian case, the ground states are generally entangled. Moreover consider the following string operators

\be
T^\m=\sum_{g_1, g_2, \cdots g_N} \chi^\m(g_1g_2\cdots g_N) |g_1, g_2, \cdots g_N\ra\la g_1, g_2, \cdots g_N|.
\ee
We may ask what is the effect of this string operator on a given ground state. In view of the relation (\ref{chichi}), it turns out that 
\be
T^\m |\Psi^\n\ra=\sum_\lambda F^{\m,\n}_\lambda |\Psi^\lambda\ra,
\ee
implying that the ground space carries a representation of the fusion algebra of the group representation. \\

\begin{figure}[t]
\begin{center}
\includegraphics[scale=0.6]{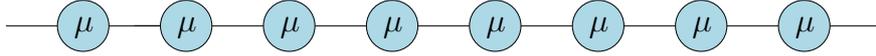}
\caption{(Color online) A ground state of the quantum Potts model is characterized by a single color (representation label) for all the sites. }
\label{gsfig}
\end{center}
\end{figure}

\subsection{Entanglement properties of the ground states}\label{entang}
We now study the entanglement properties of the ground states. First we determine how much a given site is entangled with the rest of the lattice, when the whole system is in a given ground state. Then we calculate the entanglement of a given block of length $L$. Third we determine the entanglement of two sites which are near each other or are far apart. Before proceeding to the proofs, we present the main results. When the lattice is in a given ground state $|\Psi^\m\ra$,  \\

i) the entanglement of any given site with the rest of the lattice, measured by the Von-Neumann entropy of the reduced density matrix of this site is $S(\rho^\mu_{k}) = \log(n^2_\mu)$.\\

ii) the entanglement of any given block of length $L$ is given by $
S(\rho^\mu_{[1,L]}) =\log(n^2_\mu)$ which is independent of the length of the block and verifies the area law. \\

iii) any two sites which are not nearest neighbors are separable, their reduced density matrices is a product of those of individual sites. \\

iv) any two neighboring sites are entangled and their entanglement measured by the negativity of the reduced density matrix of the two sides is given by $
{\cal N}(\rho^\m_{(1,2)})=\frac{n_\mu-1}{2}
$.\\

These results are proved in appendix A.

\section{The excited states} \label{excited states}
\begin{figure}[t]
\begin{center}
\includegraphics[scale=0.6]{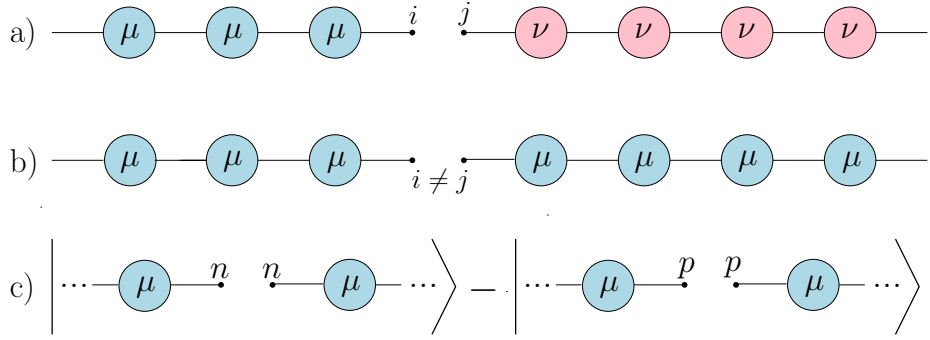}
\caption{(Color online) Different walls corresponding to different excited states, all pertaining to one unit of excitation energy: a) the two domains have different representation labels (colors). b) the two domains have the same color, but the matrix indices $i$ and $j$ are different. c)  an excited state of the kind \ref{exs3}, where both the colors and matrix indices are the same.}
\label{exs}
\end{center}
\end{figure}

Using the conventions of figure \ref{contraction} and the form of the ground state in \ref{Myground}, we can graphically depict a ground state as in figure \ref{gsfig}. This is again a generalization of the abelian Potts model, where all the sites have been colored by one single color, here a color represents a label of an irreducible representation. Each line between two bulbs means that the two adjacent indices of the matrices have been contracted (i.e. made equal and summed over). 
To obtain the excited states of \ref{hamiltonian}, consider now a cut in such a configuration as shown in figure \ref{exs}. We will see that these cuts which separate connected regions of different colors (i.e. labels of irreducible representations) produce excited states. To prove this we remind the reader that local Hamiltonians, being projectors, have eigenvalues only equal to $0$ or $1$, therefore the exited states are those in which one or more of the vertex operators have $0$ eigenvalue.  Therefore we want to find states which are common eigenstates of all the local Hamiltonians where one or more of these local  Hamiltonians have zero eigenvalues. To proceed, consider  local states like  
$|D_{m,n}^\mu\ra$.  We first obtain the action of Left and Right operators on these states, which is easily verified to be as follows: 
 \be \label{L}
 L(h)|D^\m_{m,n}\ra = D^\m_{m,p}(h^{-1})|D^\m_{p,n}\ra  \ee
 
 and
 
 \be \label{R}
  R(h)|D^\m_{m,n}\ra = |D^\m_{m,p}\ra D^\m_{p,n}(h) 
 \ee
 where summation over repeated Latin indices are implied. From these two relations we find the action of $H_i$ on the two neighboring elementary states, 
$
 |D^\m_{m,n}\ra_{i}|D^\n_{p,q}\ra_{i+1}
 $.
 Then from equations \ref{L} and \ref{R} we obtain:
 \be
 H_{i}|D^\m_{m,n}\ra_{i}|D^\n_{p,q}\ra_{i+1} = |D^\m_{m,r}\ra_{i} \left(\frac{1}{|G|}\sum_{g} D^\m_{r,n}(g)D^\n_{p,s}(g^{-1})\right)|D^\n_{s,q}\ra_{i+1}.
 \ee

Using the orthonormality relation \ref{orthonormality}, we find
  \be \label{local action}
 H_{i}|D^\m_{m,n}\ra_{i}|D^\n_{p,q}\ra_{i+1} = \frac{1}{n_\m} \delta_{n,p}\delta^{\m,\n}\sum_r|D^\m_{m,r}\ra_{i}|D^\n_{r,q}\ra_{i+1}.
\ee
Let us first consider the special case $\m\ne \n$, where we have: 
 \be \label{exs1}
 H_{i}|D^\m_{m,n}\ra_{i}|D^\n_{p,q}\ra_{i+1} =0 \hh {\rm if} \hh \mu\ne \n.
 \ee
  
 This means that if the two colors $\mu$ and $\nu$ on the two sides of the cut are different, then the local Hamiltonian $H_i$ annihilates the state and so the energy is raised by one unit. This kind of excited state is depicted in figure \ref{exs}a.

Another special case is when the two colors (i.e. representation labels) are the same, but the middle indices are different:
    \be \label{exs2}
 H_{i}|D^\m_{m,n}\ra_{i}| D^\m_{p,q}\ra_{i+1} =0 \hh {\rm if} \hh n\ne p,
  \ee
These kinds of excited states are depicted in figure \ref{exs}b. 

Finally we learn from \ref{local action} that if $\m=\n$ and $n=p$, then the right hand side of \ref{local action}, although non-zero, is independent of $n$, This means that for any two matrix indices $n, p$
     \be \label{exs3}
 H_{i}\left(|D^\m_{m,n}\ra_{i}|D^\m_{n,q}\ra_{i+1}-|D^\m_{m,p}\ra_{i}|D^\m_{p,q}\ra_{i+1}\right) =0.
  \ee 
  Hence the third kind of excitations are depicted as in figure \ref{exs}c.

The number of these new kinds of excitations for a single cut in $|\Psi^\m\ra$ which are independent is $n_\m-1$. Using the simple notation $|Q_n\ra= |D^\m_{m,n}\ra_{i}|D^\m_{n,q}\ra_{i+1}$ for brevity, they can be chosen to be $|Q_n\ra-|Q_{n+1}\ra$ ( $ n = 1, \cdots n_\m-1 $) or if we want to make them orthogonal we can choose them to be  $|Q_1\ra-|Q_2\ra$, $|Q_1\ra+|Q_2\ra-2|Q_3\ra$, $|Q_1\ra+|Q_2\ra+|Q_3\ra-3|Q_4\ra$ and so on which are obviously orthogonal. Note that if we  form their uniform summation as $\sum_n |Q_n\ra$ we will get a state  whose eigenvalue for $H_i$ is $1$, which is not an excitation any more.\\

All the excited states are derived by inserting one or more of these different kinds of cuts or excitations in a given ground state. This again generalizes a feature of the Potts model, where only one type of cut exists which separates two different values of spins or colors. In the non-Abelian case, where the representations are not one dimensional, the domain walls are more complex. They are designated by a pair of indices $i$ and $j$. This completes our analysis of the spectrum of the non-Abelian Potts model.

\section{The partition function}\label{part}
In this section we calculate the partition function which captures the statistical properties of the model. We  follow two different approaches for calculating the partition function, an algebraic  approach in which we calculate the trace of $e^{-\beta H}$ and a combinatorial approach in which we count the number of excited states of any given energy. The two approaches correspond respectively to high and low temperature expansions. The reason for this correspondence will be made clear in the derivations. 

\subsection{The algebraic approach, or  the high temperature expansion}\label{traces}
Since the local Hamiltonians commute with each other we have
\be
Z_\beta(G)=\text{Tr} e^{-\beta H} = \text{Tr} \prod_{i=1}^N e^{\beta H_i}.
\ee
Noting that the local Hamiltonians are projectors $H_i^2=H_i$, we can write the exponentials in the following linear form
\be
Z_\beta(G)=\text{Tr} \left[\prod_{i=1}^N (1+\eta H_i)\right],
\ee
where $\eta=e^{\beta}-1$. This parameter, which for high temperatures is small and for $T\lo \infty$ approaches $0$,  will act as an expansion parameter. At high temperatures the first few terms of the expansion will be a good approximation of the partition function. However in the present case, we can exactly determine the partition function by calculating all the terms. Therefore we can write 
\be
Z_\beta(G) =\sum_{n=0}^N \eta^n Z^{(n)}_\beta(G),
\ee
where $Z^{(n)}_\beta(G)$ is the trace of product of $n$ local Hamiltonians. In appendix B we prove that 

$$Z^{(k)}_\beta(G)={N\choose k}|G|^{N-k}\ \ \  k=0,\ 1, N-1 $$
and 
$Z^{(N)}_\beta(G)=K$.
Using these two results, we obtain the exact form of the partition function to be 
\be\label{ZG}
Z_{\beta}(G)= (e^{\beta}-1+|G|)^N + (e^{\beta}-1)^N(K-1).
\ee

\subsection{The combinatorial approach, or the low temperature expansion}
It is instructive to first consider the abelian Potts model. This will then teach us how to go about the non-abelian case.  In the limit  $T\lo 0$, only the ground states contribute to the partition functions. At very low temperatures both the ground states and the low lying excited states contribute to the partition function. As we will see the low lying states corresponds to few domain walls in the configuration of spins. As the temperature rises more domain walls contribute and the low temperature expansion necessitates the counting of contribution of various configurations of domain walls. Again in the present context, we can sum all the contributions exactly and obviously the result should be equal to that of the high-temperature expansion.

\begin{figure}[t]
\begin{center}
\includegraphics[scale=0.35]{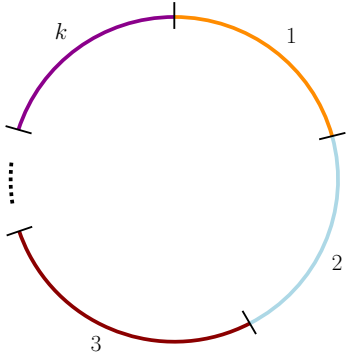}
\caption{(Color online)  An energy eigenstate of the Hamiltonian \ref{hamiltonian} (for Abelian group) is characterized by the position of the walls and different possible colors. For a configuration with $k$ fixed walls, there are $C_k(d)$ different eigenstates as given in \ref{Cd}.}
\label{walls}
\end{center}
\end{figure}

 More precisely, we note that the ground state energy of the Abelian Potts model is equal to $-N$ (when all the spins are the same, or when all the spins are colored by one single color) and the degeneracy of the ground state is equal to $d$. 
Coming to the excited states,  for each neighboring pair where the two labels $s_i$ and $s_{i+1}$  are unequal, i.e. for each domain wall, an excited state is created. So for $k$ domain walls, the energy will be $-N+k$. The degeneracy of this energy is then equal to the total number of different configurations with $k$ domain walls. This degeneracy comes from the different positions of the $k$ walls and the different possible spin configurations (which we call colors) for each configuration of walls. The first factor is easy to calculate. With a periodic boundary condition,  it is simply given by $N\choose k$. Note that $k$ should be greater than $1$, since in a periodic lattice, there cannot be one single wall. Calculation of the second factor, which we denote by $C_{k}(d)$, is done as follows. Consider figure \ref{walls} which shows $k$ different domains, separated by $k$ walls. For domain $1$ we have $d$ colors to choose. For the second domain we are left with $d-1$ walls, since the color of this domain has to be different from the first one. For the third domain we have again $d-1$ choices, since the color of this domain has to be different from the second one. So going around the circle and coming to the last domain $k$ we have $d(d-1)^k$ different choices of colors, but in so counting, we have over-counted the number of colors, since the color of the last region $k$ should be different from that of region $1$. The number of such configurations  is  simply $C_{k-1}(d)$, since in this case the two domains $k$ and $1$ merge into one single domain. Thus we arrive at the recursion relation 
 \be \label{rr}
 C_k(d) = d (d-1)^{k-1} - C_{k-1}(d), \qquad \qquad C_0(d) = d,
 \ee
 the solution of which is 
\be \label{Cd}
C_k(d)=(d-1)^k+(d-1)(-1)^k.
\ee
Therefore the degree of degeneracy of each energy level $E=-N+k$ is given by
\be
 {N\choose k}\left[(d-1)^k+(d-1)(-1)^k\right],
 \ee
leading to the partition function of the $d-$ level Potts model
\be
Z_\beta(\text{Potts}) = \sum_{k=0}^N e^{\beta(N-k)}  {N\choose k} \left[(d-1)^k+(d-1)(-1)^k\right]=(e^{\beta}-1+d)^N + (e^\beta-1)^N(d-1).
\ee

\begin{figure}[t]
\begin{center}
\includegraphics[scale=0.4]{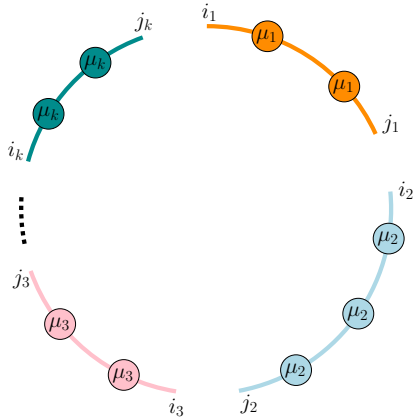}
\caption{(Color online)  An energy eigenstate of the Hamiltonian \ref{hamiltonian} (for a general finite group) is characterized by the position of the walls and the type of domains separated by them. For a configuration with $k$ fixed walls, there are $C_k(G)$ different eigenstates as given in \ref{CG}.}
\label{circle}
\end{center}
\end{figure}

We now go on to the quantum case. Here the structure or the labels of a domain wall is more complex, since as in figure \ref{circle}, each domain is characterized not only by a color (a representation label $\m$ ) but also by two matrix indices $i$ and $j$. Consider the first domain in figure \ref{circle}. The total number of choices for this  domain is thus $\sum_{\mu=1}^K\sum_{i=1,j=1}^{n_\m}=\sum_{\m=1}^K n_\m^2=|G|$. What is the number of possible choices for the next domain? From the structure of excited states in \ref{exs1}, \ref{exs2} and \ref{exs3}, we see that from all the combinations of labels for this wall, only one combination doesn't lead to an excitation at the position of the wall, hence there are $|G|-1$ possible choices for this second domain. The rest of reasoning is exactly the same as for the abelian case, namely we again have the same recursion relation as in \ref{rr}, but with a different initial condition:
\be
 C_k(G) = |G| (|G|-1)^{k-1} - C_{k-1}(G), \qquad \qquad C_0(G) = K,
 \ee
where we remind the reader that $K$ is the number of different irreducible representations or the number of conjugacy classes. This leads to 
\be \label{CG}
C_k(G) = (|G|-1)^k + (-1)^k (K-1),
\ee
and $\text{dim}(H_{-N+k})={N\choose k} \left[(|G|-1)^k + (-1)^k (K-1)\right]$ and finally to the partition function
\be\label{pG}
Z_{\beta}(G)= (e^{\beta}-1+|G|)^N + (e^{\beta}-1)^N(K-1),
\ee
in accordance with \ref{ZG}. 

In the limit $T\lo 0$, where $\beta\lo \infty$, we find from \ref{pG} that $Z_\infty(G)\lo e^{\beta N}K$, as it should be, since in this limit only the ground states contribute to the partition function and there are $K$ ground states, each with energy $-N$.  Conversely in the high temperature limit, where $\beta\lo 0$, we find $Z_0(G)=|G|^N$, meaning that all the states contribute equally to the partition function and the thermal density matrix is a completely mixed one. \\

\section{Discussion}
We have introduced the non-Abelian Potts model, and have made a rather detailed study of its properties. In the same way that Ising and Potts model have led to a large number of applications in physics, it may also be the case that the non-Abelian case, in view of its richer structure, may find such applications. On the theoretical side, we believe that there are many avenues of research which are opened by this study. Here are a few examples: \\

i) The definition of the model on 2D lattices remains to be done. Although in one dimensional lattices the local Hamiltonian \ref{localH} is the same as the one pertaining to the vertex operators of the quantum double model of Kitaev \cite{kitaev}, one cannot simply carry this to 2D lattices since it would lead to a four-body interaction, while a natural generalization of the Potts model should include two-body interactions. It may still be the case that a 2D model with two-body interaction as we have defined, is  solvable, in the sense that its ground state and low-lying states can be determined in closed form, although in 2D lattice not all the local Hamiltonians will commute with each other anymore. If this is the case, then the structure of the 2D model will be even more interesting and challenging to determine. One may even be able to establish a duality relation between the high and low temperature expansions in this case which will then enable one to determine the critical point without an exact solution.  To find an exact solution for the 2D case, the first step may be to define the model on a one-dimensional ladder, where despite the incommutability of the operators, there is still some simplicity in the geometry of the lattice. This has already been exploited to find the explicit form of the full spectrum of the Kitaev model on spin ladders \cite{Karimipour}. \\

ii) One can also generalize this to Lie groups on discrete lattices, or to Lie groups on the real line, where a suitable definition of the model Hamiltonian should be made. In view of the matrix product structure of the ground states, it may be the case that the generalization of matrix product formalism for quantum field theories as developed in \cite{ciracMPS} will be a natural framework for studying this type of generalization. \\

iii) One can add terms which pertain to an external magnetic field in the Ising model. In the same way that adding an external transverse field to the Ising model leads to a rich many-body system which undergoes a quantum phase transition, this may also happen for the non-Abelian Potts model. In such case, one anticipates that the solution will be highly non-trivial and technical, if an exact solution exists at all. Moreover the meaning of transversality of the external field should also be clarified in this case. \\

iv) It will also be interesting to study possible phase transitions between this type of order and other possibly topological orders in models where  suitable interaction terms are added to the Hamiltonian \cite{KarimipourZarkeshian, Jahromi, Hamma, Schmidt, Pascazio}. \\

v) Finally it will be interesting to investigate the relation between this 1D quantum mechanical system and the 2D classical model. Once a 2D classical possibly exactly solvable model is at hand, one can obtain the transfer matrix which represents a 1D quantum mechanical model. In the present case, knowledge of the complete spectrum of the 1D model means that the 2D classical model, once properly defined, is exactly solvable. 
These studies are now underway by the authors.\\

{}

\section{Appendix A: Proof of entanglement properties}
In this appendix we prove the results in section (\ref{entang}) on the entanglement properties of the ground states. We first  calculate the reduced density matrix of one site, which due to periodicity we take it to be the first site. 
This calculation turns out to be straightforward if we use \ref{basis} and  cast the ground state \ref{gs} into the equivalent form:

\be\label{Myground}
|\Psi^\mu\ra = \frac{1}{\sqrt{n^N_\mu}} \sum_{[ i_1 ,i_N]} |D^\mu_{i_1i_2}\ra |D^\mu_{i_2i_3}\ra \dots |D^\mu_{i_Ni_1}\ra,
\ee
where $[i_1,i_N]= \lbrace i_1,i_2,...,i_N \rbrace$, and the reduced density matrix of the first site is as follows:
\be
\rho^\mu_{1} = \text{Tr}_{\hat{1}} | \Psi^\mu \ra \la \Psi^\mu|.
\ee
Here $\text{Tr}_{\hat{1}}$ means trace over all of the spins, except the first one. Therefore we have
\be
\rho^\mu_{1} =  \frac{1}{n^N_\mu} \sum_{[ i_1,i_N], [ j_1,j_N]}   |D^\mu_{i_1i_2}\ra \la D^\mu_{j_1j_2}| \otimes \bigg( \la D^\mu_{j_2j_3}| D^\mu_{i_2i_3}\ra \otimes \dots  \la D^\mu_{j_Nj_1}|D^\mu_{i_Ni_1}\ra  \bigg).
\ee
Then using the orthonormality relation \ref{orthonormal}, we find that the terms in parenthesis give a series of Kronecker deltas as follows:
\be
 \rho^\mu_{1}= \frac{1}{n^N_\mu} \sum_{[ i_1,i_N],[ j_1,j_N]}  \delta_{i_1,j_1} \delta_{i_2,j_2} [\delta_{i_3,j_3}\dots \delta_{i_N,j_N}] |D^\mu_{i_1i_2}\ra \la D^\mu_{j_1j_2}|.
 \ee
 The sum over indices in the bracket $[\ ]$ produces a factor of $n_\m^{N-2}$ leaving us with  
 \be
  \rho^\mu_{1}= \frac{1}{n^2_\mu} \sum_{i_1,i_2} |D^\mu_{i_1i_2}\ra \la D^\mu_{i_1i_2}|.
\ee
This density matrix is diagonal in $|D^\mu_{m,n}\ra$ basis and has $n^2_\mu$ eigenvectors with $\frac{1}{n^2_\mu}$ eigenvalue and $|G|-n^2_\mu$ eigenvectors with 0 eigenvalue. Thus its Von-Neumann entropy, as a measure of bipartite entanglement between this site and the rest of the system, is equal to:
\be
S(\rho^\mu_{1}) = -\sum_n \lambda_n \log (\lambda_n)= - n^2_\mu  \frac{1}{n^2_\mu} \log(\frac{1}{n^2_\mu})=\log(n^2_\mu).
\ee

This means that only the one dimensional representations lead to product ground states and the larger the dimension of a representation, the larger will be the entanglement of the corresponding ground state. \\

To obtain bipartite entanglement between a block and the rest of the system one should obtain the reduced density matrix of a block with length $L$,  say the block consisting of sites $1$ to $L$. We denote this density matrix by $\rho^\m_{[1,L]}$ and the calculation is done as follows:  
\be
\rho^\mu_{[1,L]} = \text{Tr}_{\widehat{[1,L]}} | \Psi^\mu \ra \la \Psi^\mu|,
\ee
where $\text{Tr}_{\widehat{[1,L]}}$ means trace over all of the spins that are not in this block. Using the same arguments as above one finds that:
\bea \label{densityL}
\rho^\mu_{[1,L]} &=& \frac{1}{n^N_\mu} \sum_{[ i_1,i_N],[ j_1,j_N]} \left[  \left( |D^\mu_{i_1i_2}\ra \la D^\mu_{j_1j_2}| \otimes \dots \otimes |D^\mu_{i_L,i_{L+1}}\ra\la D^\mu_{j_L,j_{L+1}}| \right) \times \right. \\ \nonumber
&&\left. \qquad \qquad  \qquad \qquad \qquad \left( \la D^\mu_{j_{L+1},j_{L+2}}|D^\mu_{i_{L+1},i_{L+2}}\ra  \dots  \la D^\mu_{j_Nj_1}|D^\mu_{i_Ni_1}\ra  \right) \right] \\ \nonumber
&=&\frac{1}{n^N_\mu}  \sum_{[ i_1,i_N],[ j_1,j_N]}  \delta_{i_1, j_1}\delta_{i_{L+1},j_{L+1}} \left[\delta_{i_{L+2},j_{L+2}} \cdots \delta_{i_N,j_N} \right] \times \\ \nonumber
 && \qquad \qquad  \qquad \qquad \qquad |D^\mu_{i_1i_2}\ra \la D^\mu_{i_1j_2}| \otimes \dots \otimes |D^\mu_{i_Li_{L+1}}\ra \la D^\mu_{j_L,i_{L+1}}| \\ \nonumber
&=&\frac{1}{n_\mu^{L+1}} \sum_{ i_1,i_{L+1}} |\mathcal{D}^\mu_{i_1,i_{L+1}}(L) \ra \  \la \mathcal{D}^\mu_{i_1,i_{L+1}}(L)|,
\eea
where
\be
|\mathcal{D}^\mu_{i_1,i_{L+1}}(L) \ra=\sum_{i_2, i_3, \cdots i_L}|D_{i_1, i_2}\ra|D_{i_2,i_3}\ra\cdots |D_{i_L, i_{L+1}}\ra.
\ee
These states are not normalized and in fact they satisfy
$\la \mathcal{D}^\mu_{i_1,i_{L+1}}(L)|\mathcal{D}^\mu_{i_1,i_{L+1}}(L) \ra=n_\m^{L-1}$.
The reduced density matrix is diagonal in terms of these new states and has $n^2_\mu$ eigenvectors with nonzero eigenvalues equal to $\frac{1}{n^2_\mu}$  and $|G|^L-n^2_\mu$ eigenvectors with 0 eigenvalue. Therefore the Von-Neumann entropy of the reduced density matrix is equal to:
\be
S(\rho^\mu_{[1,L]}) =  - n^2_\mu  \frac{1}{n^2_\mu} \log(\frac{1}{n^2_\mu})=\log(n^2_\mu).
\ee
Therefore the entanglement of a block of length $L$ is independent of the length of the block, depending only on its two end points,  which is nothing but a manifestation of area law in this exactly solvable quantum mechanical model. \\

Finally we come to the entanglement of two different sites with each other. First we consider the entanglement of two non-adjacent sites. Following exactly the same calculation which led to the above results, it is not hard to see that the density matrix of two distant sites, say sites $1$ and $k$, will be given by
\be
\rho^\m_{(1,k)}=\frac{1}{n_\m^2}\sum_{i,j}|D^\m_{ij}\ra\la D^\m_{ij}|\otimes \frac{1}{n_\m^2}\sum_{m,n}|D^\m_{mn}\ra\la D^\m_{mn}|\equiv \rho^\m_1\otimes \rho^\m_k,
\ee 
which means that there is no entanglement between the two sites. If there is any entanglement, it is between adjacent sites. This is indeed true as we can verify. In fact we have  already found the density matrix of a block of length $L$ in \ref{densityL}. Specializing \ref{densityL} to $L=2$, we find the density matrix of two neighboring sites, say $1$ and $2$. 
\be
\rho^\m_{(1,2)}=\frac{1}{n_\mu^3} \sum_{ i,k} |\mathcal{D}^\mu_{i,k}(2) \ra \  \la \mathcal{D}^\mu_{i,k}(2)|,
\ee 
where
\be
|\mathcal{D}^\mu_{i,k}(2) \ra= \sum_{j}|D^\mu_{i, j}\ra|D^\mu_{j,k}\ra.
\ee
We can find the negativity of $\rho^\m_{(1,2)}$ by calculating its negativity in the form ${\cal N}(\rho^\m_{(1,2)})=\frac{||(\rho^\mu_{(1,2)})^{T_2}||_1-1}{2}$, where $T_2$ means partial transpose with respect to second subsystem and $||X||_1$ means the  trace-norm of $X$ which is equal to $||X||_1=\text{Tr} \sqrt{X^\dagger X}$. The partial transpose with respect to the second subsystem is equal to:
\be
(\rho^\mu_{(1,2)})^{T_2} =\frac{1}{n^3_\mu} \sum_{j,j^{\prime}} |D^\mu_{i, j} \ra \la D^\mu_{i, j^{\prime}}| \otimes |D^{\mu^*}_{j^{\prime},k} \ra \la D^{\mu^*}_{j,k} |,
\ee
where
 \be
 |D^{\mu^*}_{m,n} \ra= \sqrt{\frac{n_\m}{|G|}} \sum_g D^{\mu^*}_{m,n} (g) |g\ra.
 \ee
It is straightforward to check that $\text{Tr} \sqrt{((\rho^\mu_{(1,2)})^{T_2})^\dagger (\rho^\mu_{(1,2)})^{T_2}} = n_\m$ and the negativity of this reduced density matrix is equal to:
\be
{\cal N}(\rho^\m_{(1,2)})=\frac{n_\mu-1}{2},
\ee
which means for the higher representations of the group the ground state contains more entanglement between the nearest neighbor spins.

\section{Appendix B: Calculations of the traces in subsection \ref{traces}}
In this appendix we present details of the calculations of the partition function in the high temperature expansion. We  have
\be
Z_\beta(G) =\sum_{n=0}^N \eta^n Z^{(n)}_\beta(G),
\ee
where $Z^{(n)}_\beta(G)$ is the trace of product of $n$ local Hamiltonians. Obviously we have $Z^{(0)}_\beta(G)=\text{Tr}(I)=|G|^N$.

To calculate the first order term, consider the trace of a single local Hamiltonian like $H_1$ on the local Hilbert spaces of the first two sites $1$ and $2$, where it acts nontrivially. We write  
 \ba
\text{Tr}_{12}(H_1)&=&\frac{1}{|G|} \text{Tr}\left[\sum_{g\in G} R(g)\otimes L(g)\right]=\frac{1}{|G|}\sum_{g,h,k\in G} \la h,k| R(g)\otimes L(g)|h,k\ra\cr &=&\frac{1}{|G|}\sum_{g,h,k} \la h,k|hg^{-1}, gk\ra=\frac{1}{|G|}\sum_{g,h,k} \delta(g,e)=|G|.
\ea
This leads to $\text{Tr}(H_1)=|G|\times |G|^{N-2}=|G|^{N-1}$,  hence  $Z^{(1)}_\beta(G)=N |G|^{N-1}$. In second order we face two types of terms, those which are adjacent connected like $H_1H_2$ (which we call connected segments) or those which are far apart like $H_iH_j$ with $|i-j|\geq 2$, terms like $H_1H_3$ and so forth (which we call disconnected).
Consider a disconnected terms first. For a disconnected term like $H_1H_3$ we note that
$\text{Tr}_{1234}(H_1H_3)=\text{Tr}_{12}H_1\times \text{Tr}_{34}H_3=|G|^2$ and therefore $\text{Tr}(H_1H_3)=|G|^2\times |G|^{N-4}=|G|^{N-2}$. For a connected term like $H_1H_2$ we have  
\ba
\text{Tr}_{123}(H_1H_2)&=& \frac{1}{|G|^2}\text{Tr}\left[\sum_{g_1,g_2\in G} R(g_1)\otimes L(g_1)R(g_2)\otimes L(g_2)\right]\cr &=&\frac{1}{|G|^2}\sum_{g_1,g_2,h,k,l\in G} \la h,k,l| R(g_1)\otimes L(g_1)R(g_2)\otimes L(g_2)|h,k,l\ra\cr &=&\frac{1}{|G|^2}\sum_{g_1,g_2,h,k,l} \la h,k,l|hg_1^{-1}, g_1kg_2^{-1},g_2\ l\ra.
\ea
Therefore
\be
\text{Tr}_{123}(H_1H_2)=\frac{1}{|G|^2}\sum_{g_1,g_2,h,k,l}  \delta(g_1,e)\delta(g_2,e)=|G|,
\ee
leading to $\text{Tr}(H_1H_2H_3)=|G|\times |G|^{N-3}=|G|^{N-2}$. Therefore both types of connected and disconnected terms have the same trace which leads to a great simplification. This gives $Z^{(2)}_\beta(G)={N\choose 2}|G|^{N-2}$. This feature  is true for all the terms up to and including the  $N-1$-th term. Therefore we have $Z^{(k)}_\beta(G)={N\choose k}|G|^{N-k}\ \ \  k=0,\ 1, N-1 $. For the last terms where we have a full cycle of terms, i.e. a closed loop, the situation is a little bit different. In this case we have 
$$
\text{Tr}(H_1H_2\cdots H_N)= \frac{1}{|G|^N}\sum_{g_1, \cdots g_N}\sum_{h_1, \cdots h_N} \la g_1, g_2, \cdots g_N| h_N g_1 h_1^{-1}, h_1 g_2 h_2^{-1}, \cdots h_{N-1} g_N h_N^{-1}\ra,
$$
or

\be\label{sum}
\text{Tr}(H_1 H_2 \cdots H_N) = \frac{1}{|G|^N}\sum_{g_1, \cdots g_N}\sum_{h_1, \cdots h_N}\prod_{i=1}^N\delta(h_{i}, g_i^{-1}h_{i-1}g_i).
\ee
We can now sum over $h_N$ and then on $h_{N-1}$ down to $h_2$ and convert (\ref{sum}) to 

\be
\text{Tr}(H_1H_2\cdots H_N)=  \frac{1}{|G|^N}\sum_{g_1, \cdots g_N}\sum_{h_1}\delta(g_2 g_3\cdots g_N g_1 h_1g_1^{-1} g_N^{-1}\cdots g_2^{-1},h_1).
\ee
Using the change of variable $g_2g_3\cdots g_N g_{1}\lo g$ and performing the sum over all the other group elements $g_1$ to $g_{N-1}$, we find
where

\be
\text{Tr}(H_1 H_2 \cdots H_N)=\frac{1}{|G|}\sum_{g\in G}\sum_h \delta(ghg^{-1},h). 
\ee
The sum over $g$ gives the size of the centralizer subgroup ${\cal Z}_0(h):=\{g\in G\ \ | \  \ gh=hg\}$, which we denote by $|{\cal Z}_0(h)|$ and we remain with 

\be
\text{Tr}(H_1 H_2 \cdots H_N)=\frac{1}{|G|}\sum_{h\in G}|{\cal Z}_0(h)|=K,
\ee
where in the last equality we have used a well known identity from theory of finite groups \cite{grouptheory}.  Summing the contributions from all orders we find 

\be
Z_{\beta}(G)=\sum_{k=0}^{N-1} \eta^k {N\choose k} |G|^{N-k} + \eta^N K 
\ee
or
\be\label{ZG}
Z_{\beta}(G)= (e^{\beta}-1+|G|)^N + (e^{\beta}-1)^N(K-1).
\ee
Equation (\ref{ZG}) gives the final expression for the partition function of the non-Abelian Potts model on a ring of $N$ sites for a finite group $G$. Apart from the obvious dependence on $N$ and $\beta$, it depends on the size of the group and the number of its conjugacy classes or irreducible representations. 

\end{document}